\documentclass{article}

\usepackage{arxiv}

\usepackage[utf8]{inputenc} 
\usepackage[T1]{fontenc}    
\usepackage{hyperref}       
\usepackage{url}            
\usepackage{booktabs}       
\usepackage{amsfonts}       
\usepackage{nicefrac}       
\usepackage{microtype}      
\usepackage{lipsum}

\usepackage{amssymb}
\usepackage{color}
\usepackage{soul}
\usepackage{makecell}
\usepackage{mathtools}
\usepackage{multirow}
\usepackage{booktabs}
\usepackage{graphicx}
\usepackage{afterpage}
\usepackage{textcomp}
\usepackage{amsmath}
\usepackage{amssymb}
\usepackage{epstopdf}
\usepackage{multirow}

\title{Multi-resolution large-eddy simulation of an array of hydrokinetic turbines in a field-scale river: The Roosevelt Island Tidal Energy project in New York City}

\author{
  Saurabh Chawdhary\\
  Mathematics and Computer Science Division, \\
  Argonne National Laboratory, \\
  Argonne, Illinois 60439, USA.\\
  \texttt{sauc@anl.gov}\\
   \And
  Dionysios Angelidis \\
  Department of Civil Engineering\\
  College of Engineering and Applied Sciences\\
  Stony Brook University, Stony Brook, New York 11794, USA.\\
   \And
  Jonathan Colby \& Dean Corren\\
  Verdant Power Inc.\\
  The Octagon, 888 Main Street\\
  New York, NY 10044, USA. \\
   \And
  Lian Shen \\
  St. Anthony Falls Laboratory \\ 
  Department of Mechanical Engineering\\
  University of Minnesota\\
  2 Third Ave SE, Minneapolis, MN 55414, USA.\\
   \And
  Fotis Sotiropoulos \\
  Department of Civil Engineering\\
  College of Engineering and Applied Sciences\\
  Stony Brook University, Stony Brook, New York 11794, USA.\\
  \texttt{fotis.sotiropoulos@stonybrook.edu}\\
}

\begin{document}
\maketitle

\begin{abstract}
Marine hydrokinetic (MHK) power generation systems enable harvesting energy from waterways without the need for water impoundment. A major research challenge for numerical simulations of field-scale MHK farms stems from the large disparity in scales between the size of waterway and the energy harvesting device. We propose a large-eddy simulation (LES) framework to perform high-fidelity, multi-resolution simulations of MHK arrays in a real-life marine environment using a novel unstructured Cartesian flow solver coupled with a sharp-interface immersed boundary method. The potential of the method as a powerful engineering design tool is demonstrated by applying it to simulate a 30 turbine MHK array under development in the East River in New York City. A virtual model of the MHK power-plant is reconstructed from high-resolution bathymetry measurements in the East River and the 30 turbines placed in 10 TriFrame arrangements as designed by Verdant Power. A locally refined, near the individual turbines, background unstructured Cartesian grid enables LES across a range of geometric scales of relevance spanning approximately five orders of magnitude. The simulated flow-field is compared with a baseline LES of the flow in the East River without turbines. While velocity deficits and increased levels of turbulence kinetic energy are observed in the vicinity of the turbine wakes, away from the turbines as well as on the water surface only small increase in mean momentum is found. Therefore, our results point to the conclusion that MHK energy harvesting from large rivers is possible without a significant disruption of the river flow.
\end{abstract}


\section{Introduction}
While hydropower has been the major resource for generating electricity from water, marine and hydrokinetic (MHK) systems can harness energy from flowing natural streams of water without needing dams for impounding water \citep{khan2009hydrokinetic}.
Instead, they make use of MHK devices to extract energy from rivers, tidal channels, and ocean currents.
Axial hydrokinetic turbines are a popular choice of device to extract MHK energy 
and have been deployed in the field \citep{torrens2017assessment, williamson2016self, web:openei-axial-1} as well as studied analytically, experimentally and computationally \citep{nishino2012efficiency, batten2008prediction, gant2008modelling, kang2012reallife, kang2014onset, malki2014planning, chawdhary2017wake, angeloudis2018optimising}.

Despite the recent advances in numerical methods and the exponential growth of computing power, numerical study of MHK turbine arrays at field-scale rivers remains a major challenge. 
Marine environments are dominated by complex bathymetry and a wide range of natural and man-made structures that give rise to turbulent flows dominated by energetic, large-scale coherent structures. Introducing MHK devices in such a complex fluid flow environment gives rise to additional complexities in the turbulent flow due to turbine-turbine and turbine-bathymetry interactions, which give rise to flow phenomena spanning a broad range of temporal and spatial scales. Such phenomena need to be understood and quantified in order to optimize MHK arrays for energy extraction and evaluate the structural integrity and reliability of the devices in the harsh waterway fluvial environment \citep{li2011integrated, kang2012numerical,tatum2016wave,musa2018performance}.
Computational modeling presents the only viable approach for gaining such understanding at field scale and on a site-specific basis.

Computational tools capable of tackling such complex flows, however, need to be able to resolve waterway and device induced turbulence, which occurs at high Reynolds numbers, across a range of scales, and is dominated by inherently 3D energetic coherent structures. 
For the most part, field-scale simulations of large rivers have been performed using two-dimensional (2D) models by solving the shallow-water equations \citep{abderrezzak2009modelling, casulli2000unstructured, heniche2000two, yoon2004finite, lee2017development} or statistically stationary 3D models based on the Reynolds-averaged Navier-Stokes (RANS) equations \citep{ge20053d,nagata2005three,lu20093d,baranya2015flow}, due to their simplicity and computational expedience. 
Computational methods for carrying out eddy-resolving simulations in real-life rivers have only recently began to appear in the literature \citep{kang2015numerical, khosronejad2016high, khosronejad2016large, wilcox2017simulation}. Such models, however, have yet to tackle the simulation of MHK devices in real-life river environments.

\citet{james2010simulating} used a modification of the Environmental Fluid Dynamics Code (EFDC) \citep{hamrick2007environmental} developed at Sandia National Laboratories (SNL) to simulate changes in the marine environment caused by an array of MHK turbines, by modeling MHK turbines as single energy extraction points in a 2D domain.
Other previous studies have used unsteady Reynolds-Averaged Navier Stokes (RANS) modeling to simulate flows in idealized channel geometries with actuator disc model to model the turbines, but without accounting for the prevailing features of the surrounding marine environment \citep{bai2009investigation, abolghasemi2016simulating}. 
LES has been widely applied as a powerful computational tool in the highly relevant field of wind farm optimization \citep{churchfield2012large, yang2013large, porte2013numerical, vasel2017wind, yang2018large}. Conclusions from such studies, however, are of limited use for MHK energy systems due to inherent differences between the ambient atmospheric flow in wind farms (e.g. the size of the wind turbine relative to the thickness of the atmospheric boundary layer) and the fully developed turbulence in waterways (where the MHK turbines occupy a considerable part of the flow depth) \citep{chamorro2013three}. \citet{churchfield2013large} are developing a framework for simulating MHK turbine arrays in natural waterways \citep{churchfield2013large} 
and have applied it to study the effects of turbulence in the incoming flow on the wake characteristics in an artificial straight channel using LES and actuator disk parameterization.
\citet{wilcox2017simulation} performed site-specific detached-eddy simulation (DES) of turbulent flow in the Fundy Tidal Region at field scale to gain insight about the placement of tidal turbines. However, the simulations with turbines was planned for future work.

Sotiropoulos and co-workers have recently developed an advanced computational fluid dynamics framework, the Virtual Flow Simulator (VFS-Rivers), capable of carrying out site-specific LES in real-life waterways with stationary and/or mobile river beds \citep{sotiropoulos2015hydraulics}. VFS-Rivers employs the curvilinear immersed boundary (CURVIB) method \citep{ge2007numerical, kang2011high} with wall models for carrying out LES in arbitrarily complex domains with arbitrarily complex immersed boundaries and migrating bedforms. The predictive capabilities of VFS-Rivers have been extensively demonstrated for flows in man-made field scale streams \cite{kang2011high}, natural streams \citep{khosronejad2016large}, and large rivers under baseline and extreme flooding conditions \cite{khosronejad2016high}. VFS-Rivers has also been applied to carry out high-fidelity LES of a single MHK turbine \citep{kang2012reallife, kang2014onset} and MHK turbines in a TriFrame arrangement \citep{chawdhary2017wake} in a laboratory straight open channel. In all cases, excellent agreement with laboratory and field-scale (where available) measurements have been reported and the ability of the code to uncover new physics that could not be accessed by experiments alone has been demonstrated (see for example \citet{kang2014onset}).

 \begin{figure}[htp]
  \center
  \includegraphics[width=5.7in, keepaspectratio=true]{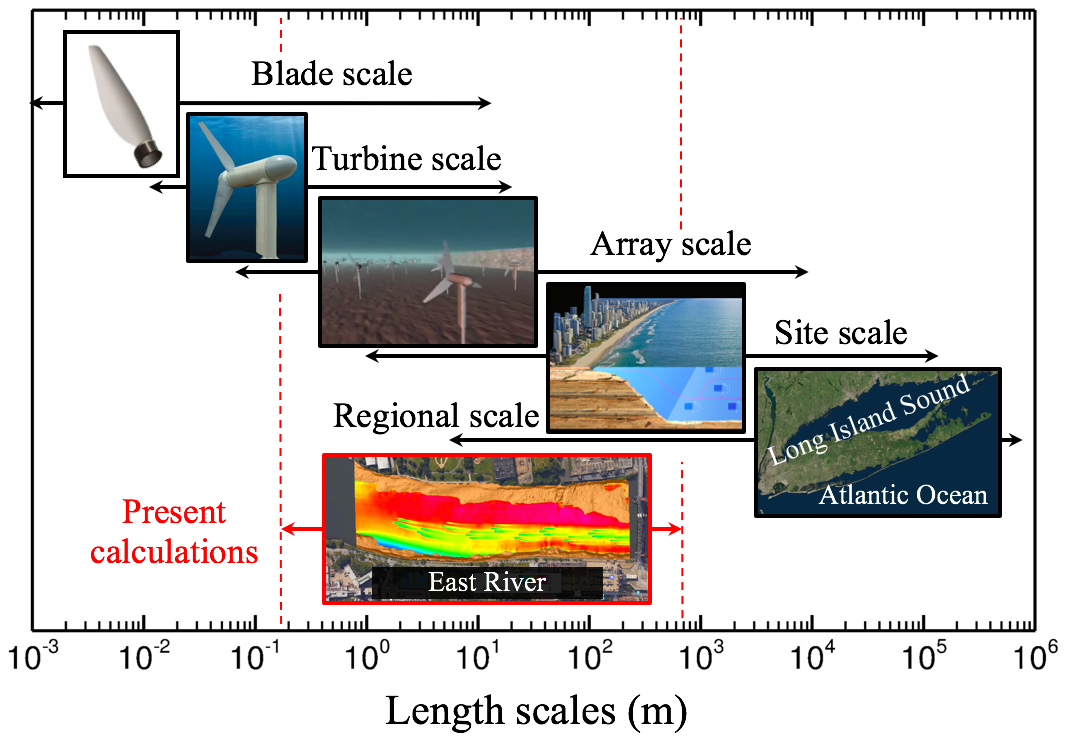}
  \caption{Schematic showing the range of length scales present in hydrodynamic modeling of MHK energy systems at different physical dimensions involved. (Adapted from \citet{adcock2015tidal}). The simulation in this paper resolves flow scales across multiple ranges, as marked in red.}
  \label{fig:scales-plot}
\end{figure}
 
In spite of the aforementioned recent progress, high-fidelity computational methods that can carry out LES of MHK turbines and multi-turbine arrays in real-life waterways, resolving waterway and device-induced unsteady coherent structures, have yet to be proposed in the literature. The reason for this is illustrated in Fig. \ref{fig:scales-plot}, which describes the various length scales present in the flow when a turbine is placed in a natural marine environment like a tidal channel. The blades of the turbine can be very thin giving rise to small scale flow structures (of the order of centimeters) whereas the turbine diameter itself is normally of the order of few meters. A multi-turbine array introduces flow scales which, depending on the array configuration, can be beyond an order of magnitude larger than the turbine diameter. Furthermore, when modeling a particular array installation site or a larger marine environment, the flow characteristics of the region (river and/or ocean system) also come into play as they determine the ambient and inflow conditions to the turbine array.
As recent as few years ago, \citet{stoesser2010calculation} noted that ``because of the
 high computational cost, LES can currently only be used for lower Reynolds numbers
 making it not directly applicable in the river engineering practice where usually Reynolds numbers are above $10^6$ and computational domains sizes range from several hundred meters  to kilometers''. 
Other researchers have also acknowledged the computational challenges in dealing with natural geometries \citep{abad2008flow, kang2012numerical}.
\citet{kang2012reallife, kang2014onset, gant2008modelling, batten2008prediction, malki2014planning} and \citet{chawdhary2017wake} have studied flow past turbines using LES to resolve scales in one or two of the scale ranges shown in Fig. \ref{fig:scales-plot}.

In this work we seek to demonstrate for the first time that LES of multi-turbine MHK farms in real-life waterways resolving coherent structures spanning the scales of the river to the scale of the turbine rotor and wake are now well within reach. For that, we implement a recently developed numerical method capable of local mesh refinement on unstructured Cartesian grids \citep{angelidis2016unstructured} in the new version of the VFS-Rivers code to enable multi-resolution simulations of MHK arrays in a real-life waterway. The specific case we consider herein is the Roosevelt Island Tidal Energy (RITE) project currently under development in the East River in New York City by Verdant Power. We employ high-resolution bathymetry measurements and the turbine layout proposed by Verdant Power to reconstruct a virtual model of the East River tidal power plant. The resulting digital model of the East River power plant is embedded in a background Cartesian mesh locally refined in the vicinity of the turbines to carry out LES at field scale using the method developed by \citet{angelidis2016unstructured}. 
Our work makes a number of novel contributions to the literature by: a) Demonstrating the potential of the geometry-based adaptive mesh refinement methodology in performing high-fidelity, multi-resolution LES for real life aquatic environments; b) Bridging computationally for the first time the large disparity of scales between the coherent structures induced by the energy harvesting device and those induced by the waterway within which the devices are embedded in so that turbine-turbine and turbine-waterway interactions can be studied numerically at field scale and c) illustrating the utility of site-specific LES as a powerful tool for enabling simulation-based optimization of MHK powerplants at field scale.

This paper is organized as follows. In section \ref{sec:method}, we briefly discuss the numerical method used for the simulations. In section \ref{sec:validation}, we validate the numerical method by simulating a model wind farm and comparing simulations with measurements. Section \ref{sec:eastriverSetup} describes the deployment site of the hydrokinetic turbine array simulated herein. Section \ref{sec:baseline} presents results from baseline simulations of the relevant reach of the East River without the turbine array. Section \ref{sec:array-simulation} presents simulation results for the 30 turbine array installed in the East River and discusses its effects on the river flowfield. Finally, in section \ref{sec:conclusion} we summarize the findings of this work and discuss future directions.


\section{Numerical Methods}
\label{sec:method}
The new version of VFS-Rivers, incorporating the unstructured Cartesian flow solver of \citet{angelidis2016unstructured}, is employed herein to perform multi-resolution LES past the RITE turbine array.
The equations governing the instantaneous, resolved flow field for three-dimensional, incompressible, turbulent flow are the spatially-filtered continuity and Navier-Stokes equations employed in LES models. The sub-grid scale (SGS) stress tensor representing the effects of the SGS motions on the resolved fields of the LES is modeled using the Smagorinsky model \citep{smagorinsky1963general}. The governing equations are discretized on a $3D$ hybrid staggered/non-staggered grid layout \citep{Ge20071782} adapted on unstructured Cartesian grids using the second-order central differencing scheme when stencil cells are at same level of refinement. 
When computational cells are surrounded by other cells of varying level of refinement, hanging nodes and a bi-quadratic operator  are utilized to facilitate the spatial discretization using a hybrid second-order upwind scheme.
Time integration of governing equations is performed using the fractional step algorithm such that the discrete divergence of the velocity field is satisfied to machine zero at every time step. 
The implicit solution of the momentum equation is achieved by using a Jacobian-free Newton-Krylov method coupled with the generalized minimal residual solver (GMRES) of \citet{saad1986gmres}.
The unstructured flow solver is efficiently parallelized using the portable, extensible toolkit for scientific computation (PETSc) libraries and message passing interface (MPI) \citep{petsc-efficient, petsc-web-page, petsc-user-ref}. The PETSc libraries are also used to solve the momentum equation using the aforementioned Jacobian-free Newton-Krylov method. The high performance preconditioners (HYPRE) libraries \citep{falgout2006numerical} with BoomerAMG \citep{yang2002boomeramg} preconditioner are used for solving the Poisson equation.
We use the actuator line model for modeling the axial turbines in the flow where the blades of turbines are represented using rotating line forces in the flow \citep{angelidis2016unstructured, yang2015large}. 
The sharp interface immersed boundary method approach in VFS-Rivers is employed to simulate arbitrarily complex immersed boundary in the flow domain -- the river bathymetry in this case. In this method, the immersed irregular structures are represented using unstructured triangulated surface meshes embedded within a locally refined Cartesian background grid. 
The effect of immersed boundaries on the flow (i.e. the river bathymetry in this case) is accounted for by reconstructing at every time step boundary conditions at the nodes in the immediate vicinity of the boundary via interpolation along the local normal to the boundary direction. Given the high Reynolds number of the flows of interest, the interpolation to reconstruct velocity boundary conditions is implemented using a wall model (see \citet{yang2012computationalWT} for details).
The freesurface boundary of the flow channel is treated as a rigid lid, an assumption which has been successfully applied in the previous works of \citet{kang2011high, kang2012reallife, kang2014onset} and \citet{yangtoward} among others, for low Froude number flows ($Fr \approx 0.2$ for the flow in the East River).
More details about numerical method can be found in \citet{angelidis2016unstructured}.


\section{Model validation : Flow over aligned wind turbine array}
\label{sec:validation}
%
%
In \citet{angelidis2016unstructured} we validated the unstructured Cartesian flow solver for several cases including turbulent flows over stand-alone wind and hydrokinetic turbines. 
In this section we seek to demonstrate the predictive ability of the solver in LES of turbine arrays. Since no data is available for MHK arrays, we employ herein the results of a well documented wind tunnel experiment for a model wind farm.

\subsection{Test case and computational details}
We carry out LES of turbulent flow past a 6$\times$3 aligned array of model wind turbines spaced $5D$ and $4D$ apart of each other in the streamwise (z) and spanwise (x) directions, respectively. Each model wind turbine has diameter $D=0.15m$ and hub height $h=5D/6$. The tip-speed ratio (TSR), $\lambda$, for all the turbines is $\lambda=4.1$ and the Reynolds number, $Re$, based on the rotor diameter, $D$, and the mean incoming hub height velocity, $W_h$, is $Re=2.5\times10^4$.
Details of the corresponding wind tunnel experiment can be found in the work of \citet{chamorro2011turbulent_b}. We note that the same test case has also been investigated numerically by \citet{yang2013large} using the Virtual Flow Simulator (VFS) \citep{VFS} to carry out LES on a structured Cartesian grid without refinement.

The computational domain extends 12$D$, 5$D$ and 33$D$ in the spanwise ($x$), vertical ($y$) and streamwise ($z$) directions, respectively. Two simulations are carried out, the first is performed on a uniform structured Cartesian grid (G1) with $N_x$=121, $N_y$=61 and $N_z$=331 gridnodes in $x$, $y$ and $z$ directions, respectively, and the second is performed on a locally refined mesh (G2) which is generated after remeshing the primary structured Cartesian grid (G1) in a cylindrical manner around the mid-column rotors, by applying up to two levels of refinement. Thus, G1 consists of $2.4\times10^6$ cells, and the resulting grid spacing around all the turbine rotors is $D/10$, $D/12$ and $D/10$ in $x$, $y$ and $z$ directions, respectively. To generate the unstructured grid G2, we apply two levels of refinement around the mid-column wind turbines in a cylindrical manner and the center of the finite cylinders is located at the turbines' hub height. The first level of refinement is located within a cylindrical region with a radius of $R_{ref1}=0.7D$ which extends between $0.8D\le z \le32D$ from the inlet and the second level of refinement is applied within six cylindrical zones with a radius of $R_{ref2}=0.6D$ and length of $L_{ref2}=2D$ located symmetrically around the turbine rotors.
The equivalent uniform structured Cartesian grid having the resolution same as that around the turbine rotors, would consist of $1.5 \times 10^8$ computational cells. 
The resolution  around the turbine rotors of the middle column in G2 is equivalent to the finest spatial discretization used in the work of \citet{yang2013large} using a stretched Cartesian structured grid with $17.3 \times 10^6 $ cells. 
To generate inflow conditions, we carry out LES for a domain of the same cross-section as the computational domain with the turbines but without the turbines and with periodic conditions applied in the streamwise direction. This simulation is continued until statistical convergence is achieved. Instantaneous flowfields from the statistically stationary flowfields are stored and subsequently used to prescribe instantaneous boundary conditions at the inlet of the wind farm domain.
Free slip conditions are applied at the top and spanwise boundaries and the wall model of \citet{1063/1.1476668} is used as boundary condition on the bottom boundary. 
The turbines are parameterized using the actuator line model as described in \citet{yang2015large}.
The time step is equal to $\tau=0.001D/W_h$. The simulations are first run for 25 revolutions of the turbines until the total kinetic energy of the domain reaches a quasi-steady state. Subsequently, the results are time averaged for another 55 rotor revolutions to compute the statistics presented below.

\subsection{Comparison with experimental measurements and numerical data}
\begin{figure}[!t] 
  \centering
  \includegraphics[width=0.75\linewidth]{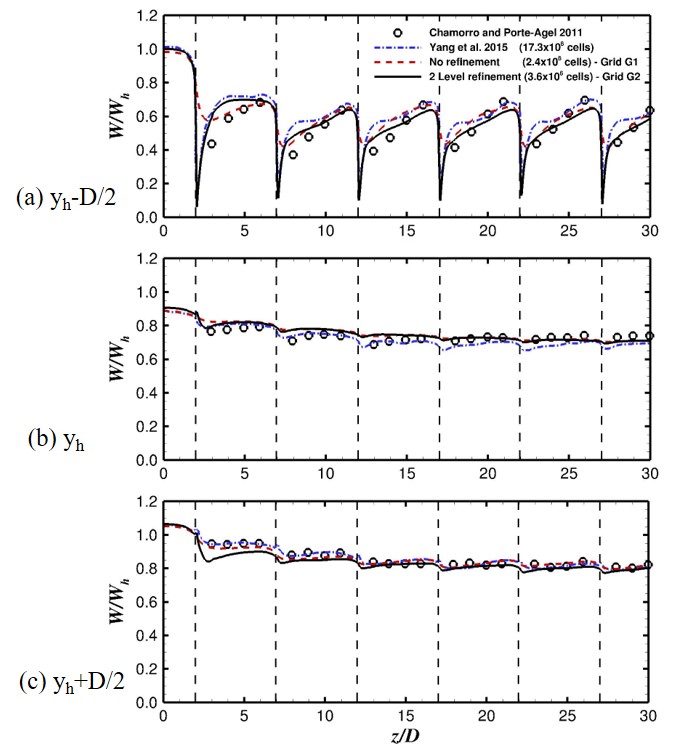}
  \caption{
Comparison of the mean streamwise wind velocity at different vertical locations, along the streamwise direction, on a $x$-$z$ plane passing through the center of the middle column wind turbines. Calculations performed on a uniform Cartesian grid $121\times61\times331$ without refinement (G1) and on a 2-level refined grid adapted around and in the wake of the wind turbines' rotor (G2). Comparison is performed against previous experimental measurements of \citet{chamorro2011turbulent_b} and numerical simulations of \citet{yang2013large}.
  }
  \label{stremwiseveldistr} 
\end{figure}

Streamwise profiles of the calculated (with and without refinement) time-averaged streamwise velocity (W) are compared with the measurements of \citet{chamorro2011turbulent_b} and the LES of \citet{yang2013large} at different vertical locations on a $x$-$z$ plane passing through the center of the middle column wind turbines in Fig. \ref{stremwiseveldistr}. At the bottom tip of the wind turbines (Fig. \ref{stremwiseveldistr}(a)), the present calculations agree well with the experimental measurements with a small over-prediction of the time-averaged streamwise velocity observed in the wake of the first and second wind turbines. Our results are in very good agreement with the experimental measurements in the wake of the fourth, fifth and sixth wind turbines where the calculations of \citet{yang2013large} slightly under-predict the streamwise velocity. Overall the calculations performed on a 2-level locally refined grid agree very well with the experimental measurements. It is seen in Fig. \ref{stremwiseveldistr}(b) that the calculations performed on a locally refined grid, G2, result in better prediction of the near-wake time-averaged streamwise velocity compared to calculations performed on the primary grid G1 and calculations on stretched Cartesian grids \citep{yang2013large}. At the top tip of the wind turbines (Fig. \ref{stremwiseveldistr}(c)), small discrepancies are observed mostly in the wake of the first wind turbine; however, very good agreement is obtained for all the other locations. The maximum discrepancy found 1D downstream of the first turbine is approximately 10\%. The maximum discrepancy in the wake of the rest of the turbines at hub height is less than 5\% (at a location approximately 1D downstream of the second turbine). 

\begin{figure}[!b]
\centering
\includegraphics[width=\textwidth]{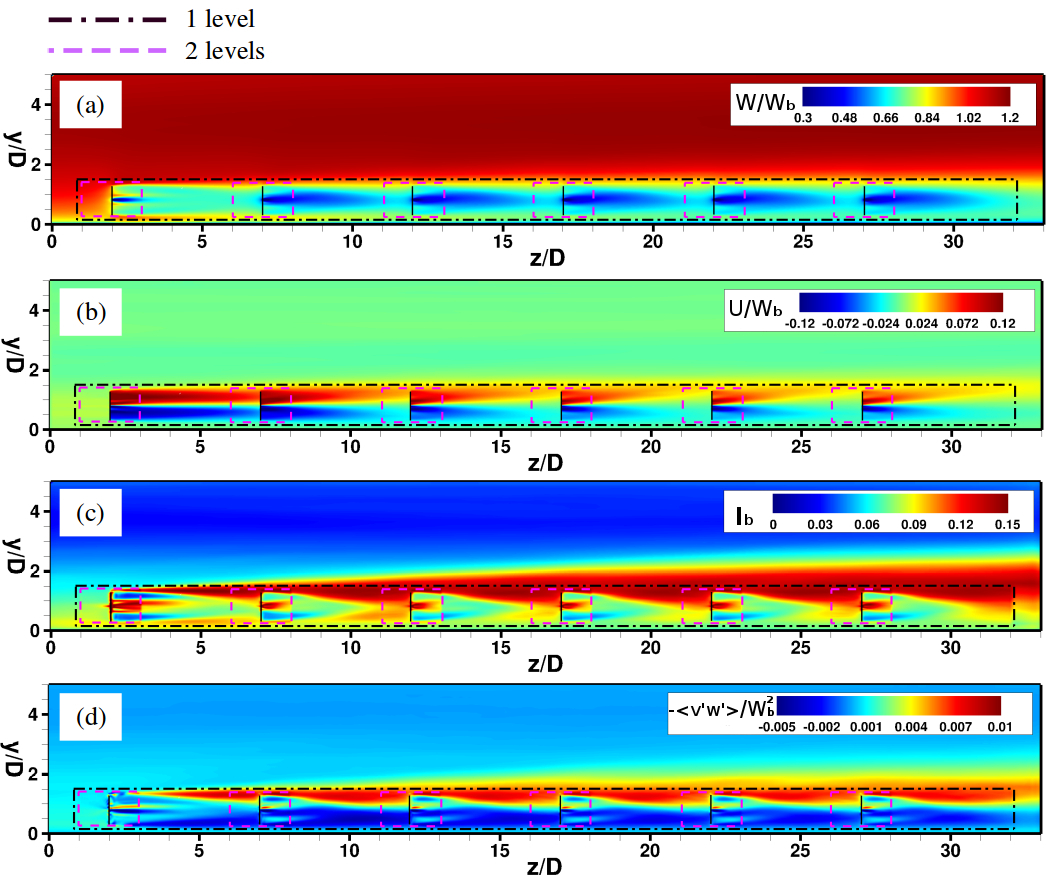}
\caption{
Contours of the time-averaged: (a) streamwise wind velocity; (b) spanwise wind velocity; (c) streamwise turbulence intensity; (d) Reynolds shear stress, on a $y$-$z$ plane passing through the center of the wind turbines located in the middle column. The plotted values are normalized by the incoming hub height velocity, $W_h$, and the calculations are performed on a 2-level refined grid consisted of $3.6\times10^6$ computational cells.
}
\label{Fig_time_averaged_contours_2}
\end{figure}

Figure \ref{Fig_time_averaged_contours_2} shows contours of the mean velocity and turbulence statistics along the vertical plane passing through the center of the wind farm. In line with the observations of \citet{yang2013large}, the velocity deficit in the wake of the first wind turbine is much weaker compared to the wakes of the other turbines which are nearly the same (Fig. \ref{Fig_time_averaged_contours_2}(a)). The actuator line model can also predict the significant levels of time-averaged spanwise velocity induced by the rotating turbines; which is more intense in the wake of the first and second turbines (Fig. \ref{Fig_time_averaged_contours_2}(b)). The distribution of the time-averaged velocity profiles provides evidence of the smooth transition of the flow properties across cells with different levels of refinement \citep{angelidis2016unstructured}. Figure \ref{Fig_time_averaged_contours_2}(c) depicts contours of the streamwise turbulence intensity computed on a 2-level refined grid. The velocity fluctuations in the wake and at the top tip of the wind turbines result in the formation of a layer of enhanced turbulence intensity, which transitions to higher levels with downwind distance. 
The turbulence intensity is larger at the top tip compared to the bottom tip shear layer as a result of the interaction of the turbine wakes with the wall boundary layer. We may also observe the development of a region with significant turbulence intensity just behind the turbines at the hub height. Figure \ref{Fig_time_averaged_contours_2}(d) shows the contour plots of Reynolds shear stress, $-<v'w'>$, normalized by the incoming hub height velocity, $W_h$. The turbines introduce positive kinematic shear stress in the wakes at the top-tip and negative stress below the turbine hub height, respectively. The absolute kinematic shear stress at the top tip height is significantly larger compared to the values in lower regions, indicating that the high levels of vertical flux of streamwise momentum are associated with the intense shear layer. Overall, the results presented in this section show that our calculations are in good agreement with the experimental measurements of \citet{chamorro2011turbulent_b}. The discretization of the governing equations on a locally refined unstructured Cartesian grid enabled achievement of comparable level of accuracy with the calculations of \citet{yang2013large} while using 5 times smaller number of computational cells.

\section{The East River site and computational setup}
\label{sec:eastriverSetup}
The East River is a tidal strait in New York City that connects New York Harbor and the Atlantic Ocean to Long Island Sound. An aerial view of New York City in Fig. \ref{fig:ManhattanEastRiver}(a) shows its location between Manhattan and Queens.
The present study is part of the RITE Project, a pilot project which aims to install up to one Megawatt of MHK power in the East Channel of the East River. 
The Federal Energy Regulatory Commission (FERC) has issued a commercial pilot license (\#P-12611) to \textit{Verdant Power}. The region where MHK turbine array is proposed to be installed is indicated in Fig. \ref{fig:ManhattanEastRiver}(c).
The East River being a tidal strait starts and ends in ocean and the flow is driven by the difference in tidal heights with a semi-diurnal time period of about 12 hours.
In spite of the varying nature of the flow, it is possible to treat the flow as unidirectional for the purpose of our study for the following reasons: 
(1) the time scale (12 hours) of flow pulsation is much larger than the time taken for the hydrodynamics within the simulated section to attain a statistically converged steady state -- the simulations were run for a total physical time of $~900\,s$;
and (2) the turbines in the flow are allowed to yaw so that they are approximately aligned with the primary flow direction.
The tidal channel bed is a cobble layer of rocks larger than the size that can be mobilized by the flow eliminating the need for sediment transport modeling.

\subsection{Deployment of the Triframes in East River}
The map in Fig. \ref{fig:ManhattanEastRiver}(c) shows the FERC licensed boundary within which the turbines are to be installed. 
Verdant Power has developed Kinetic Hydro-Power System (KHPS) turbines that will be used to harvest kinetic energy from the channel. These are 3-blade axial turbines $5\,m$ in diameter.
The power generation capacity of these machines, which naturally varies with time as a function of the instantaneous water flow velocity, is $56\;kW$ and their rated power in the RITE site flow velocity distribution is $35\,kW$ \citep{vpireportv2}.
Verdant Power has successfully performed deployment of individual turbines at the site.
The proposed plan is to deploy up to 30 turbines in the East Channel of the East River in several phases with constant testing and environmental monitoring. 
For structural reasons and ease of deployment the turbines are to be installed on the river bed in 10 TriFrame arrangements -- i.e. sets of three turbines mounted on the apexes of a common triangular base. The proposed locations of the TriFrames are marked in Fig. \ref{fig:ManhattanEastRiver}(c) by triangles.
Underwater cables from each unit connect to the onshore controls and the electric grid.
The estimated average annual production from the array is between $1680-2400\,MWh$ \citep{vpireportv2}.
The electricity produced will be added to the grid and used to power buildings in Roosevelt Island.

\subsection{Construction of the digital terrain model}
\begin{figure}[!b]
\centering
\includegraphics[width=160mm]{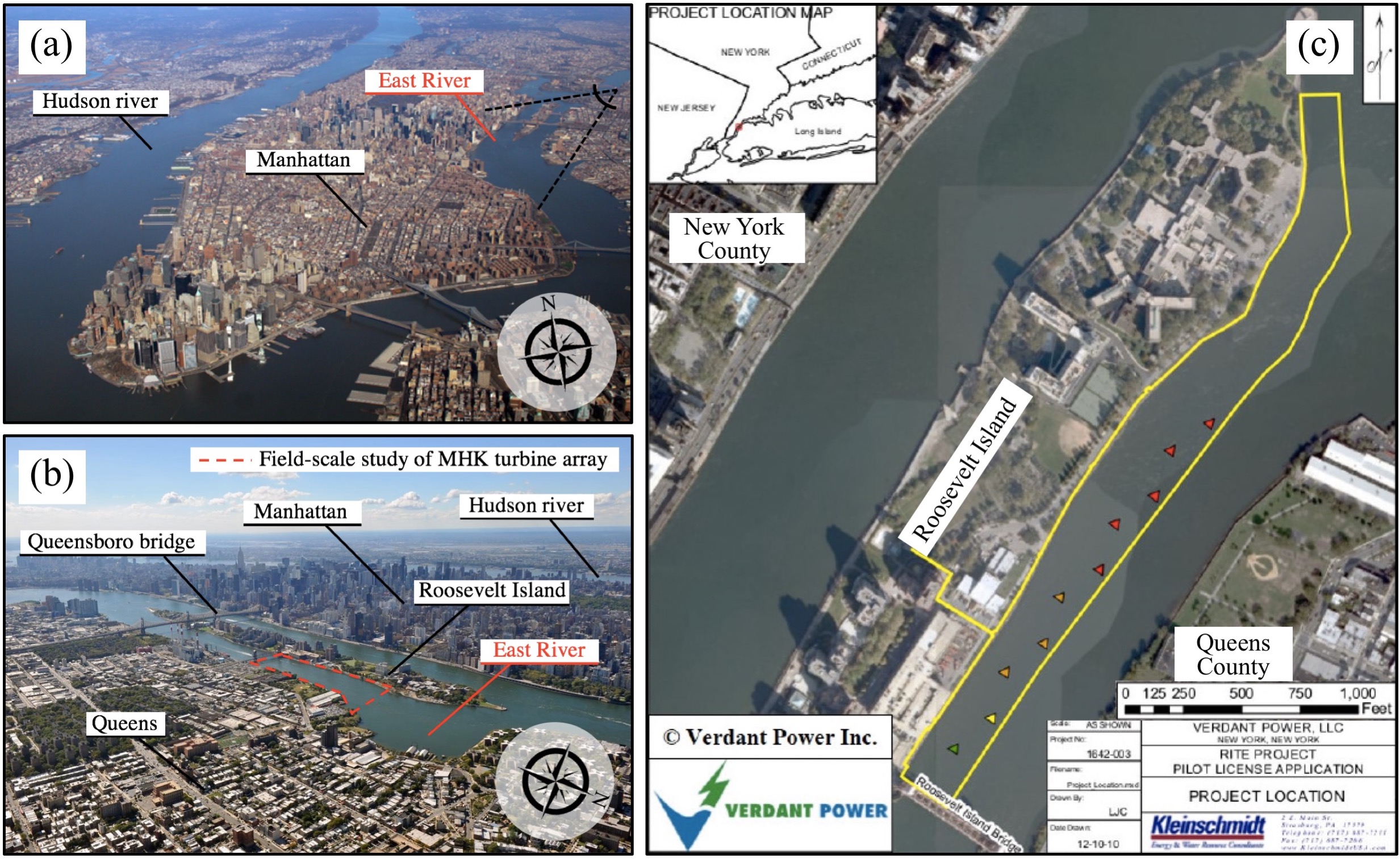}
\caption{Perspective view of the New York City area: (a) North-northeast (NNE) panoramic view of Manhattan; (b) West-southwest (WSW) view of the East River and showing the location of the field-scale study of marine hydrokinetic turbine array, map data \textcopyright 2017 Google. (c) RITE project boundaries with proposed TriFrame locations ($\Delta$). Copyright of Verdant Power Inc. \citep{vpireportv2}.}
\label{fig:ManhattanEastRiver}
\end{figure}

High-resolution bathymetry data were collected within the deployment site for the RITE project (see Fig. \ref{fig:ManhattanEastRiver}(c)) in 2015 on a grid of $0.15\,m \times 0.15\,m$.
Using this bathymetry survey, a digital elevation model of the river was created by discretizing the river bed with an unstructured triangular mesh with approximately $2.2\times 10^5$ triangular elements and $4.4\times 10^5$ points. 
Some regions within the river, where survey data was missing, were truncated and/or interpolated to provide the bathymetry immersed boundary.
The digital elevation model for the $675\,m$ long simulated reach is shown in Fig \ref{fig:eastriver-domain}(a).
The width of the channel in this reach varies between $175\,m$ and $220\,m$ while 
the flow depth varies between $2\,m$ and $13\,m$ with most common depth over the reach being approximately $10\,m$. Depths are relative to NAVD 88 datum.

\section{East River Simulation Without Turbines}
\label{sec:baseline}
In this section we report the results of LES of the turbulent flow in the East River reach without the turbines at the nominal inflow value. 
This baseline simulation will provide the baseline of the river flow to subsequently investigate the effect of installing the array of turbines in the river.
The simulation results are also compared with limited field measurements in terms of a single vertical velocity profile measured using Acoustic Doppler Current Profiler (ADCP).


\subsection{Computational setup}
\label{sec:baseline-setup}
\begin{figure}[htp]
  \center
  \includegraphics[width=\textwidth, keepaspectratio=true]{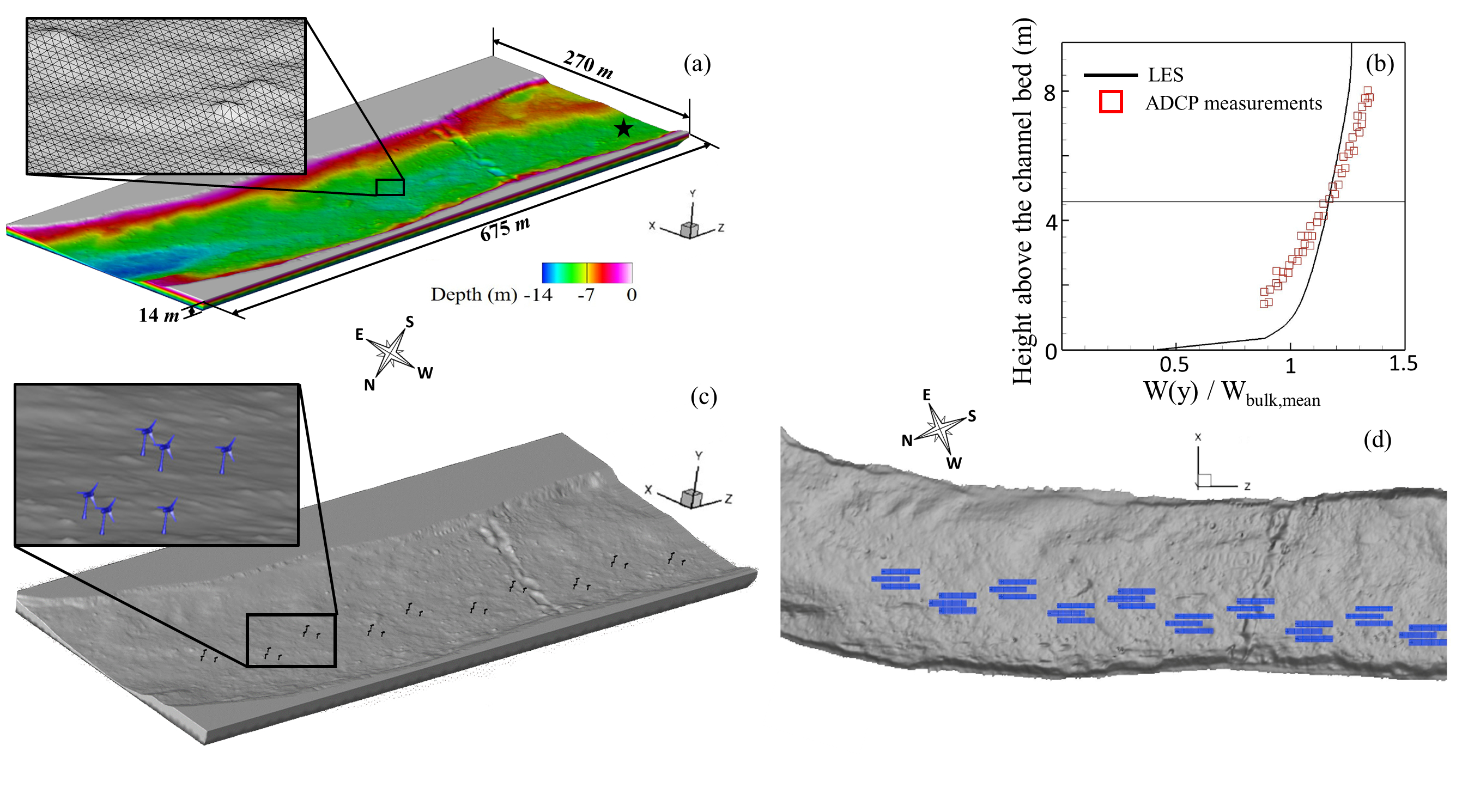}  
  \caption{Computational domain and digital elevation model of simulated channel. A sample channel bathymetry is shown in the inset window represented by unstructured mesh. A star in (a) marks the location where an Acoustic Doppler Current Profiler (ADCP) was placed for measurement in the field.
(b) Comparison of computed and measured velocity in the East River. Horizontal solid line denotes hub-height position if a turbine were to be placed here.
(c) East River model with array of 30 turbines placed in TriFrame arrangement and (d) local grid refinement in the wake of each turbine.}
  \label{fig:eastriver-domain}
\end{figure}

A Cartesian box of size $720\,m$, $14\,m$ and $270\,m$ in the streamwise (Z), spanwise (X) and vertical (Y) directions, respectively, was chosen as the computational domain within which the digitally reconstructed river bathymetry is immersed (see Fig. \ref{fig:eastriver-domain}).
The shore-line for the river was extracted using the ArcGIS mapping tool from a satellite map made available within the tool. 
This ensured that the physical shore-lines of the river are accurately captured in the computational domain.
The bathymetry was then extended outside of the river boundaries to span the computational domain in the XZ plane (see Fig. \ref{fig:eastriver-domain}(b)). 
Finite positive depth ($1\,m$) was assigned to on-shore region in the land.
The outlet boundary section was artificially extended by repeating the last bathymetry cross section over $100\,m$ in the streamwise direction. 
This artificially created straight section of the river is used to facilitate application of outflow boundary conditions and eliminate spurious reflections at the downstream boundary that could contaminate the simulated flow and destabilize the flow solver. 

Fig. \ref{fig:eastriver-domain}(a) shows a sample of the unstructured triangular mesh which has been generated to discretize the river's bed topography, as needed by the sharp interface immersed boundary method in VFS-Rivers.
The bottom bed was treated as a rigid bed with no sediment transport since it is made of large rocks on bedrock. 
A wall modeling approach, as described earlier, was used for the boundary condition on the bed.
At the inlet boundary of the domain, instantaneous flowfields from a pre-computed fully developed turbulent flow are prescribed so that statistically the inflow corresponds to fully developed turbulent flow through a straight channel with cross-section identical to the most upstream cross-section of the East River bathymetry used in the simulations. That is, the so prescribed flowfields were obtained from a precursory LES through a straight channel with periodic boundary conditions in the streamwise direction. 
Naturally, the so prescribed inflow conditions do not correspond to the actual state of the river flow at the entrance of the East River reach we simulate herein.
However, we note here that for the turbine simulations presented next, the streamwise distance between the inflow section and the most upstream turbine in the array is 12D allowing for sufficient distance for the flow to develop over the actual river bathymetry before it encounters the turbines.
At the downstream outlet boundary of the domain, Neumann (zero streamwise gradient) boundary conditions were imposed for all three velocity components. The river free surface is treated as rigid lid.

The nominal flow rate in the river and representative velocity based on the ADCP-measurements by Verdant Power were approximated to be $3.28\times10^3\,m^3/s$ with a $2\,m/s$, respectively.
From the bathymetry survey, the modal depth in the channel is approximately $H=10\,m$. 
Using the above values of velocity and depth, the Reynolds number in the river is $2\times 10^7$, which is typical value for river flows.
The background Cartesian domain was discretized with $360\times 40\times 1036$ cells in the X, Y and Z directions, respectively, i.e. a total of $1.49\times 10^7$ grid cells. Grid points are distributed uniformly in all three directions, 
resulting in spacing of $0.75\,m$, $0.35\,m$ and $0.7\,m$ in the X, Y and Z directions, respectively. The time step in the calculations corresponds to Courant-Friedrichs-Lewy (CFL) number of $CFL=0.03$.
The simulations were carried out until the total kinetic energy in the domain is statistically converged. Subsequently, the flow-field was averaged for approximately 3 flow-through times where a flow-through time is defined as time taken to travel the streamwise extent of the domain at velocity $W_b$, which is the bulk inlet velocity.
The simulations were performed on 264 processors of HPC clusters composed of Intel 2.4 GHz processors for approximately 1 month.

\subsection{Comparisons with field measurements}
\label{sec:baseline-results}
A single velocity profile was measured in the river using ADCP along the depth at the Universal Transverse Mercator (UTM) coordinates of (588987.03 E, 4513118.35 N) in zone 18N, hereafter known as ADCP-N location also shown in Fig. \ref{fig:eastriver-domain}(a).
Figure \ref{fig:eastriver-domain}(b) shows comparison of computed mean velocity with the field measurements in the East River provided by Verdant Power. 
Any discrepancies between the experimental measurements and the numerical calculations could be attributed to the following.
The river bathymetry used in this simulation was obtained in 2015 whereas the ADCP velocity measurements at the above-mentioned location were obtained nearly 2 years earlier. This time difference naturally resulted in discrepancies in the exact bathymetry in the vicinity of ADCP-N location due to natural evolution of the river bed thereby altering the flowfield. 
The reported flow-depth from the measurements was $9.16\,m$ approximately 10 \% less than $9.82\,m$ observed during the 2015 survey.
Moreover, the exact flow-rate of the river at the time of the measurements was unknown and it was only approximated using the point measurement and assuming same mean velocity across a cross-section approximately perpendicular to the nominal flow direction in the channel. 
Due to the nature of the site, it is extremely difficult to obtain high fidelity field measurements and turbulence statistics that are accurately correlated with the time of tide in the East Channel.
Hence, in the limit of available data and considering the variability of the processes involved, we argue that the simulations are in general agreement with the measurements.
The average relative error is around 6\%, which is acceptable considering the complexity of the calculation and the uncertainties mentioned above.
We also point to the extensive validation studies of the VFS-Rivers code for flows in rivers and streams and for cases for which detailed mean flow and turbulence statistics measurements were available \citep{kang2011high, khosronejad2016large}, which build confidence in the accuracy of the present simulations.
More results for this case are shown in next section in comparison with the turbine array simulation.

\section{Simulation of East River with the turbine array}
\label{sec:array-simulation}
In this section, we carry out and discuss the results of LES of 10 TriFrames of axial hydrokinetic turbines installed on the river bed per the array design developed by Verdant Power is performed.
Using the Verdent Power's Generation 5 Free Flow System axial hydrokinetic turbines \citep{FreeFlow2018}, the 30 turbine array is placed in the East River section as shown in Fig. \ref{fig:eastriver-domain}(c).
The turbine rotors are $5\,m$ in diameter and the hub height is $4.58\,m$ from the bottom.
This height is fixed due to the construction of the TriFrame and is uniform for all turbines. 
The distance between the turbine's hub from the free surface is variable depending on the depth of the channel at the individual turbine installation location.
The geometrical details of the TriFrames as well as their hydrodynamic efficiency and wake characteristics have been the subject of a recently published paper by \citet{chawdhary2017wake}.
The TriFrames from first to last, in the positive Z direction, are numbered as TF-10 to TF-1 in reverse order in the Verdant's proposed location.

The computational grid for this simulation is derived from the one used for the baseline case in the previous section.
Two-level geometry-based adaptive mesh refinement was applied to the baseline grid to obtain high resolution in the wake region of the turbines.
The local refinement was along an elliptical cylinder with the central axis of the cylinder aligned with the center of the turbines. 
For each turbine, the extent of the refinement region is 1.4D and 2D in vertical and spanwise directions, respectively.
The refinement region further extends from 1D upstream of turbine location to 7D downstream of the turbine.
The grid refinement procedure added approximately $9.4 \times 10^6$ cells, resulting in the post-refinement grid having $2.43\times 10^7$ number of Cartesian cells.
The grid cells were Cartesian and aspect ratio was 2.1:1:2 for all the cells in X, Y and Z directions, respectively.
The local resolution in the turbine wake was $0.1875\,m$, $0.0875\,m$ ,$0.1872\,m$ in the X, Y and Z directions, respectively, which corresponds to 27, 57, 29 points per turbine diameter.
Such a grid resolution is known to give acceptable results for the turbine modeled with an actuator line model \citep{yangtoward}. Actuator line models have been previously used successfully to predict the far wake dynamics of axial hydrokinetic turbines \citep{kang2014onset, churchfield2013large} and wind turbines \citep{yang2013large, porte2013numerical, angelidis2016unstructured}.
Also, it is evident from prior numerical studies \citep{angelidis2016unstructured} and the results we present herein that the proposed numerical framework allows turbine simulations on locally refined grids without introducing any artificial reflection of the vortical structures at the coarse/fine interface of the computational mesh.

The rest of the flow and boundary conditions were the same as used for the baseline flow simulation in the Section \ref{sec:baseline-setup}.
The simulation was run until the total kinetic energy in the domain is statistically converged which happened in approximately 1.2 flow-through times to obtain turbulent statistics. 
The flow-field was then averaged for another 5 flow-through times.

\subsection{Results}
\begin{figure}[htp]
  \center
  \includegraphics[width=\textwidth, keepaspectratio=true]{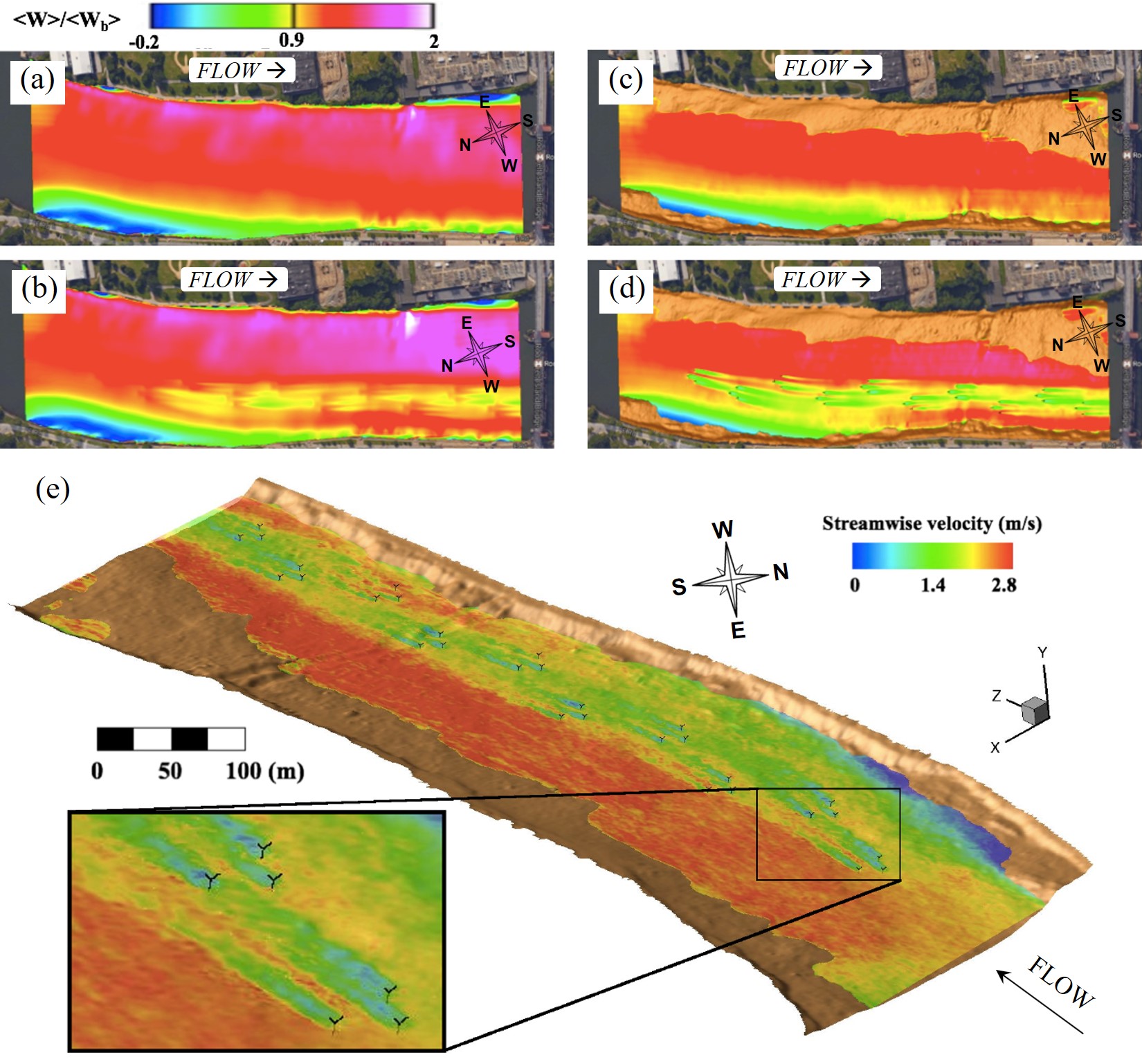} 
  \caption{Contours of time-averaged streamwise velocity of the East Channel of East River, normalized by the bulk mean inflow velocity, in an XZ plane at free surface 
(a) without turbines, (b) with turbines, and at $7\,m$ depth 
(c) without turbines, (d) with turbines.
(e) Contours of instantaneous streamwise velocity in an XZ plane $7\,m$ below the free surface with TriFrames.
}
  \label{fig:eastriver-W}
\end{figure}

Figure \ref{fig:eastriver-W}(a) and (b) shows the contours of time-averaged streamwise velocity at the freesurface of the channel. The results from the baseline simulation without turbines are also shown using the same contour levels for comparison.
The baseline simulation shows that the velocity at the freesurface is higher in the upper half South-East side of the river. 
The kink in the geometry North-East of the East River, which is essentially upstream of the simulated section of the channel, is responsible for the generation of a significant wake region in the flow. Hence, the velocity distribution is greatly effected by the strong deceleration in northwest of the east channel of the East River.
Its effect on the simulated section is caused because of the precursory simulation for generating the inflow. 
The domain for inflow generation was generated by extrusion in the streamwise (Z) direction.
This is a reasonable approximation because the channel upstream of the inflow section is straight at the end of curvature. As we already discussed, this was necessitated because bathymetric data was available only for a limited channel reach.
Ideally, the simulated channel reach should extend upstream of the present channel and VFS-Rivers is capable of handling such an extended domain. However, realizing this requires more detailed bathymetry data and is beyond the scope of the present work.
The contours from the simulation with the turbine array installed show a similar trend as the baseline case with high velocity in the upper half of the river.
There is a weak signature of the turbine wakes marked by the region of reduced velocity on the free-surface of the channel. The footprint of the turbine wake on the surface flow is more pronounced for the TriFrames in the middle of the array. The first two TriFrames barely show any effect on the free surface. This is because the hub height of these turbines is much lower compared to the other TriFrames -- this is evident from the digital elevation model showing depth contour in the river (Fig. \ref{fig:eastriver-domain}(a)).

In Fig. \ref{fig:eastriver-W}(c) and (d), the contours of time-averaged velocity are shown on a horizontal plane $7\,m$ below the free surface. 
The hub of the upstream turbines is located approximately $7\,m$ below the free surface.
In Fig. \ref{fig:eastriver-W}(c), we note for the baseline simulation a distinct feature of the bathymetry on the bottom-left corner where a ridge is seen to contribute to the slow velocity wake in this region discussed earlier.
In Fig. \ref{fig:eastriver-W}(d), the wakes of the turbines are clearly visible. 
The wakes of the first three TriFrames, specifically the first and second, do not align with the axis of rotation. This indicates that the yawing mechanism of the turbines should turn them in the primary incoming flow direction which, for these turbines, is not in Z directions. 
The yawing mechanism was not modeled in the present case (instead the turbines were assumed fixed in direction), hence we see the wake alignment departing from the axis of rotation.
The overall wake of the array suggests that the staggering of the 10 TriFrames as  proposed by Verdant Power is more efficient for energy extraction because of higher incoming flow momentum for the downstream TriFrames.
Additionally, Fig. \ref{fig:eastriver-W}(e) shows contours of instantaneous streamwise velocity on a plane $7\,m$ below the free surface, i.e. near the hub-height of the first TriFrame. This contour plot clearly shows the complex dynamics of the flow captured in the simulation. 
The large range of the scales present in the model can be readily seen. 
The $5\,m$-diameter turbine is orders of magnitudes smaller than the dimensions of the river. Yet, the unstructured Cartesian grid enabled us to locally refine the grid and resolve the flow in the wake of turbines.
The complex interaction of the scales resolved by the refined grid is shown in the inset figure for a smaller highlighted region.
An equivalent uniform structured grid, with resolution equal to the maximum achieved in the wake of the turbines, would consist of a staggering $9.5\times10^8$ cells (close to one billion); making such a ``brute-force" calculation extremely demanding from the computational standpoint.
Being able to carry out LES across such a broad range of spatial scales using relatively coarse grids (compared to those that would be required for ``brute-force" simulations) is only possible using the proposed multi-resolution modeling framework with local refinement.

\begin{figure}[htp]
\center
  \includegraphics[width=\textwidth, keepaspectratio=true]{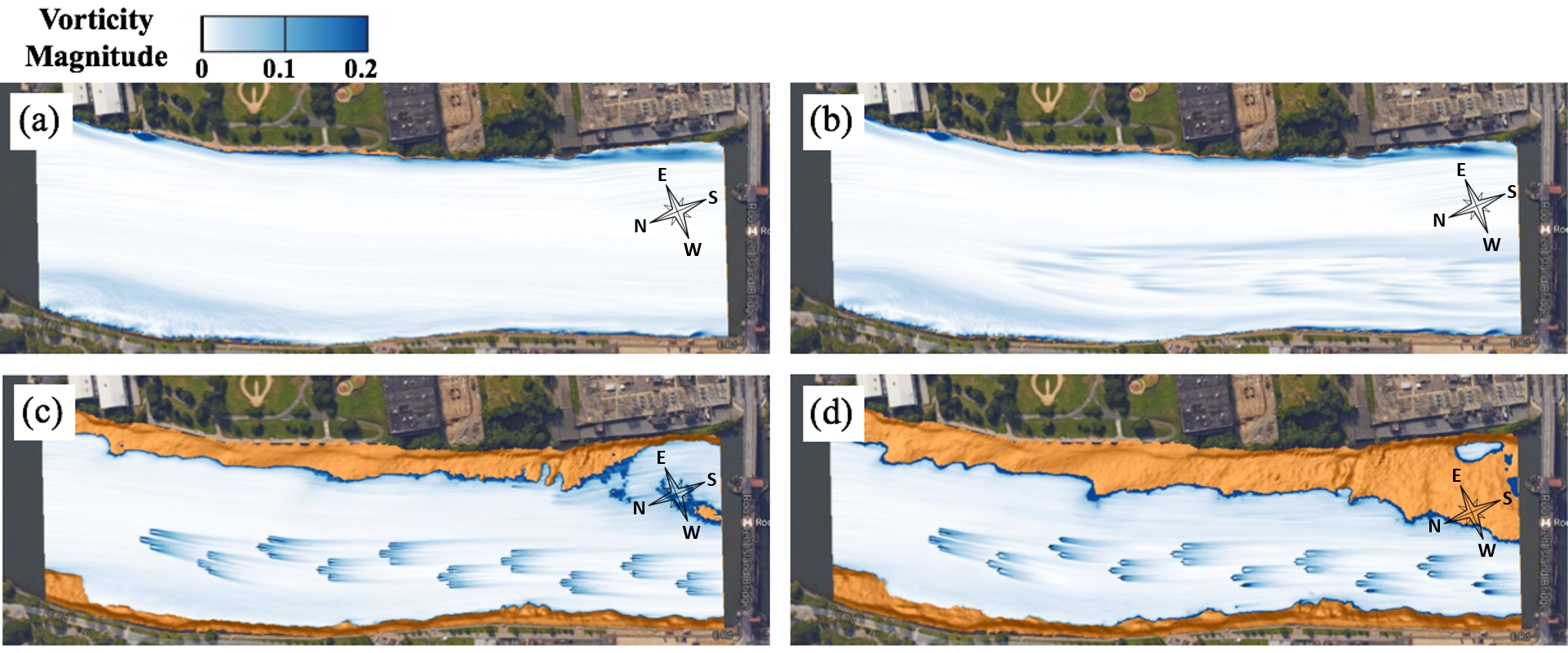} 
  \caption{Contours of time-averaged vorticity magnitude normalized by the bulk mean inflow velocity $W_b/D$ and rotor diameter $D$ in an XZ plane of East River (a) without turbine array at free surface, and with 10 TriFrames of turbines installed at (b) free surface, (c) $5\,m$ depth and (d) $7\,m$ depth.}
  \label{fig:eastriver-vortmag}
\end{figure}

Figure \ref{fig:eastriver-vortmag}(a) shows contours of time-averaged vorticity magnitude at the free surface of the river without the turbine array. There is no significant vorticity at the free surface for the scale shown here, except for the vorticity generated by the river banks.
For the simulation with the turbine array, Fig. \ref{fig:eastriver-vortmag}(b) shows that the TriFrames make only a weak mark on free surface vorticity values. In fact their surface signature is only visible using the exaggerated scales selected for this purpose in Fig. \ref{fig:eastriver-vortmag}(b).
In Fig. \ref{fig:eastriver-vortmag}(c) and (d), the time-averaged vorticity contours are shown at the XZ planes $5\,m$ and $7\,m$ below the free surface of the channel, respectively. 
The wakes of the turbines show higher vorticity magnitude, which was absent when no turbines were installed.
The vorticity magnitude contours follow the low velocity wakes similar to the streamwise vorticity distribution. 
Higher vorticity also exists near the banks and other geometrical features of the river bed as seen in Fig. \ref{fig:eastriver-vortmag}(c) near the downstream end of the channel.

\subsection{How TriFrames affect the flow in East River}
 \begin{figure}[th]
  \center
  \includegraphics[width=310pt, keepaspectratio=true]{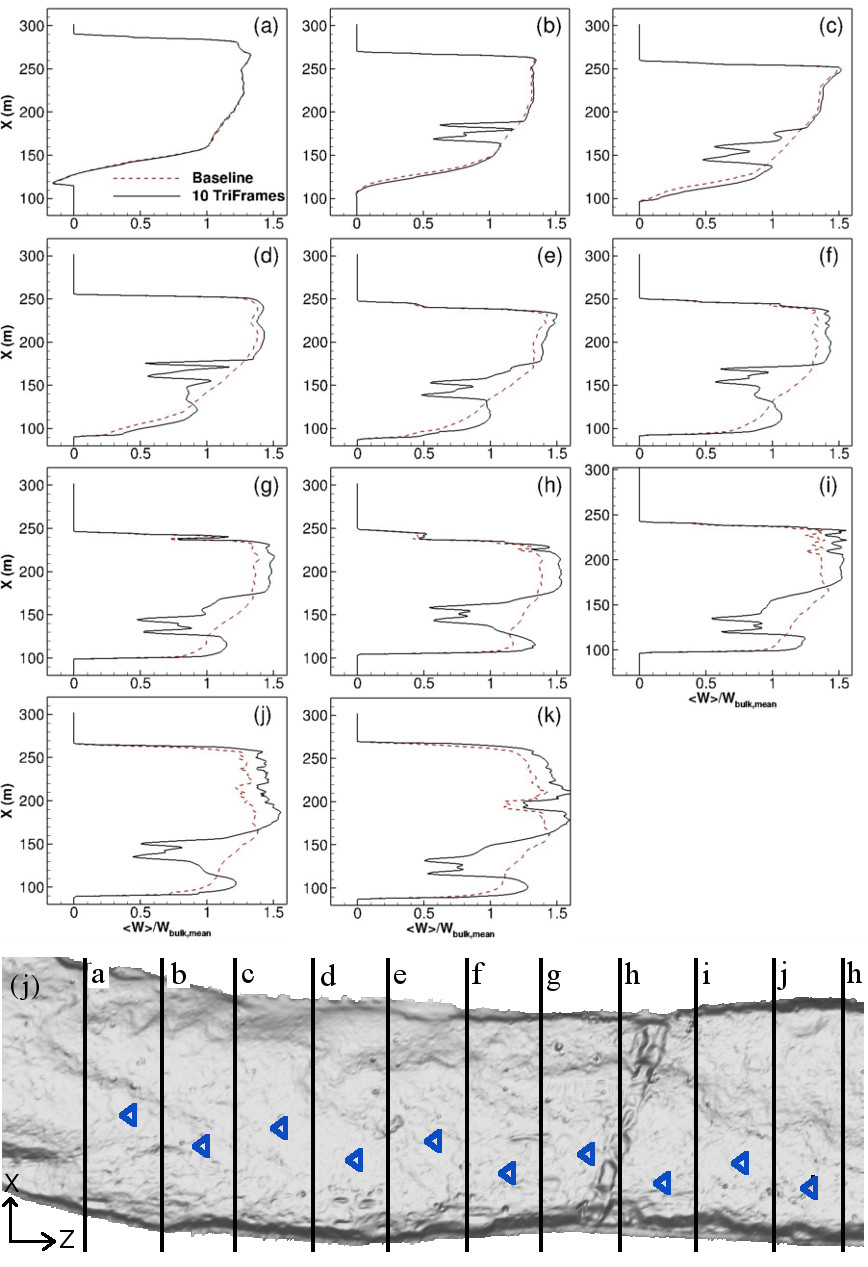} 
  \caption{Transverse profiles of streamwise velocity on the XZ plane $5\,m$ below the free surface of the East River along spanwise (X) direction with and without 10 TriFrames of turbines installed at locations (a)$Z=-23.8\,m$ (b)$Z=36.2\,m$ (c)$Z=93.6\,m$ (d)$Z=154.4\,m$ (e)$Z=213.2\,m$ (f)$Z=274.5\,m$ (g)$Z=332.7\,m$ (h)$Z=393.9\,m$ (i)$Z=454.1\,m$ (j)$Z=515.2\,m$ (k)$Z=569.3\,m$. Location (a) is 6D upstream of first TriFrame and following locations are 6D downstream of successive TriFrames. (l) shows the lines along which profiles were extracted.}
  \label{fig:Wprofiles}
\end{figure}

The presence of the turbine array alters the flow patterns in the river by extracting momentum from the stream flow and increasing the turbulence kinetic energy (TKE). In Fig. \ref{fig:Wprofiles} we compare the calculated mean streamwise velocity profiles at several downstream location in the river, approximately halfway between the TriFrames. 
In the first plot \ref{fig:Wprofiles}(a), the velocity 6D upstream of the first TriFrame is the same for the baseline as well as flow with the turbines. 
A possible reason for this could be the effect of not accounting for the blockage effect the turbines would impart on the large-scale East River flow whereby the flow could bypass to the west channel of the East River if a larger region was considered in the simulation. However, we stipulate that given the small blockage ratio of a TriFrame (0.033\%) the bypass flow to the West Channel of the East River would not be significant \citep{chen2011blockage,nishino2012effects} -- the validity of this assertion, however, remains to be confirmed by future simulations considering a much large section of the East River with the Roosevelt Island included in the model.
In the next plot Fig. \ref{fig:Wprofiles}(b), 6D downstream of the first TriFrame, flow de-acceleration is observed in the spanwise location of the turbines. The W-shaped profile marks the signature of the two downstream turbines of the TriFrame. This shape is similar to the TriFrame wake signature from a single wake TriFrame study. 
The profile in Fig. \ref{fig:Wprofiles}(c) appears to be more similar to the TriFrame wake signature since the hub of second TriFrame is closer to $5\,m$ depth.
Similar wake signature is seen for the following TriFrames. In Fig. \ref{fig:Wprofiles}(d) through \ref{fig:Wprofiles}(k), the presence of turbines result in higher flow velocity in the bypass region around the turbines. Noticeably, the span where there was no turbine shows only slightly (up to 10\% in \ref{fig:Wprofiles}(i)) higher streamwise velocity.

\begin{figure}[th]
  \center
  \includegraphics[width=410pt, keepaspectratio=true]{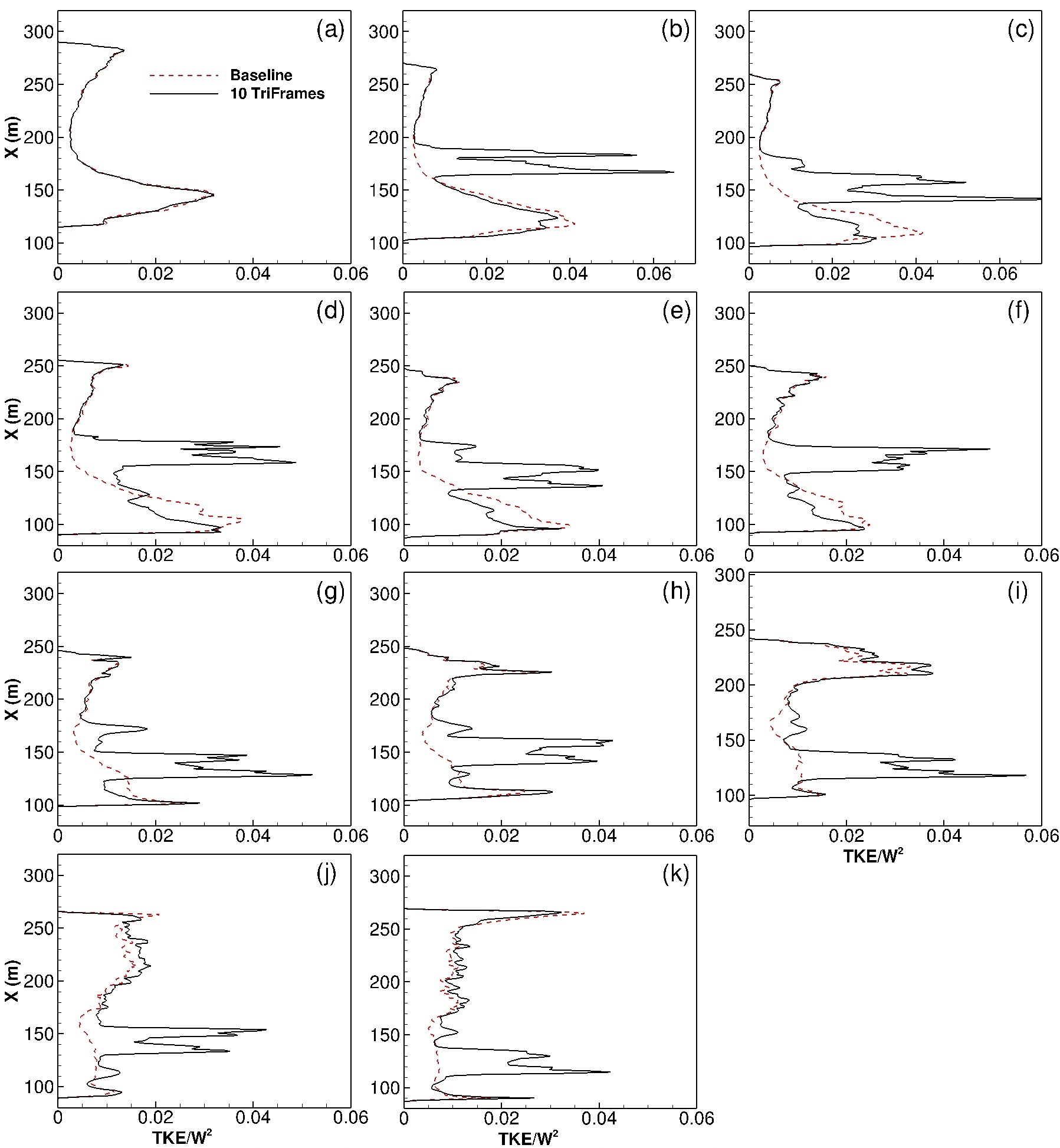} 
  \caption{Turbulent kinetic energy (TKE) profiles in XZ plane $5\,m$ below the free surface of the East River along spanwise (X) direction with and without 10 TriFrames of turbines installed at locations (a)$Z=-23.8\,m$ (b)$Z=36.2\,m$ (c)$Z=93.6\,m$ (d)$Z=154.4\,m$ (e)$Z=213.2\,m$ (f)$Z=274.5\,m$ (g)$Z=332.7\,m$ (h)$Z=393.9\,m$ (i)$Z=454.1\,m$ (j)$Z=515.2\,m$ (k)$Z=569.3\,m$. Location (a) is 6D upstream of first TriFrame and following locations are 6D downstream of successive TriFrames. Fig.\ref{fig:Wprofiles} (l) shows the lines along which profiles were extracted.}
  \label{fig:TKEprofiles}
\end{figure}

Fig. \ref{fig:TKEprofiles} shows transverse profiles of the TKE at the same locations as in Fig. \ref{fig:Wprofiles} with and without the turbine array installed.
As expected, the incoming flow upstream of the array shows relatively higher levels of TKE generated by the river banks. This is consistent with the incoming flow fed as the inflow boundary condition which was generated by assuming a straight channel extending upstream of the river with the same cross-section as the inlet section.
The remaining profiles in Fig. \ref{fig:TKEprofiles} (b through k) reveal higher levels of TKE in the wake of every TriFrame, which is the expected effect of TriFrame added turbulence in the baseline river flow. It is worth noting, however, that the effect of the TriFrames is restricted in the immediate vicinity of the turbines as the levels of turbulence near the left (relative to the turbines as shown in Fig. \ref{fig:Wprofiles}) bank are almost identical with and without turbines.

\section{Conclusions}
\label{sec:conclusion}
In this work we demonstrated the use of a powerful numerical framework to successfully perform a LES at site-scale of a natural marine environment with a MHK turbine array. Such numerical undertaking was possible with the use of high performance computing and the new generation of the VFS-Rivers code featuring local mesh refinement on unstructured Cartesian grids. 
Locally refined grids allowed the bridging of scales between the river channel and the turbines and the wakes they induce on a grid with significantly smaller (by two orders of magnitude) number of grid nodes than that would be required using a ``brute-force" approach--i.e. discretizing the entire river channel with a uniform grid with spacing small enough to resolve the wake of each turbine.
More specifically, we were able to perform simulations on a grid with 24 million grid nodes with local resolution near the turbines ranging from D/27 to D/57. Similar resolution with the brute-force approach would require close to one billion grid nodes.
Simulations were first performed in the river for a nominal flow rate without the turbines to obtain baseline flow characteristics.
Field measurements in the river were made using an ADCP to obtain a vertical profile of streamwise velocity at a single location.
The velocity profile from the baseline flow was compared with this available velocity profile. 
The prediction of velocity profile showed reasonable agreement with the field data but also underscored the difficulties in obtaining measurements of sufficient quality in harsh tidal environments to enable detailed validation of computational methods at field scale. Future work in this site should focus on obtaining simultaneous measurements of river bathymetry and flow statistics at several locations through the channel to enable a more comprehensive validation of the flow solver for the baseline case.

LES of the East River channel with the 30 turbine array in place, arranged in 10 TriFrames, provided valuable insights about the hydrodynamics of the array and its impact on the baseline river flow.
The bathymetry of the river site resulted in distinct site-specific flow patterns that can only be simulated by data-informed simulations such as those we reported herein.
Our simulations indicated a marginal acceleration on the baseline river flow in the regions where turbines were not present. Comparison with the baseline flow in terms of mean streamwise velocity as well as vorticity magnitude indicates that there is a very small signature of the turbine wake at the free surface of the channel, thus, suggesting that the array would have essentially no effect on the navigability of the East River while it operates. Overall, the surface effects of the array are found to be negligible compared to the free surface disturbances already present in the baseline flow of tidal channel.
The wakes of the first few TriFrames of the array were not aligned with the axis of rotation defined by orientation of turbines.
This means that the yaw of the turbines should turn the turbine in the direction of the incoming flow. 
The yaw mechanism of the turbine was not modeled in the present simulation.

Future work will focus on refining the model to eliminate uncertainties so that we can obtain better answers regarding the site specific performance of the proposed array and the deployment site.
In this study, for instance, only limited bathymetry data was available upstream of the turbine array. A more extensive bathymetry of the site is needed in order to avoid the unwanted effects created by the inflow boundary treatment due to the missing upstream section.
Geometry-based adaptive grid refinement was adopted in the present simulations. However, in the future, temporal adaptive mesh refinement may also be adopted in the simulations, similar to many astronomical simulations \citep{iapichino2008hydrodynamical, fryxell2000flash}. Coupled with efficient repartitioning algorithms, such simulations will enable dynamic resolution of coherent structures in the wake of the turbines while optimizing the computational efficiency.
Such capability will be especially useful when the yaw mechanism of the turbines will be include din the simulations to enable dynamic wake stirring.

\section{Acknowledgement*}
\label{sec:acknowledgments}
This work was  supported by National Science Foundation (NSF) grant number \textit{IIP-1318201}. 
The bathymetry survey for the simulated section was provided by Verdant Power Inc. through a contractor Ocean Survey Inc. Computational resources were provided by the Minnesota Supercomputing Institute (MSI) at University of Minnesota.


\bibliographystyle{apalike}

\end{document}